\documentclass[aps,prl,preprint,groupedaddress]{revtex4}
\usepackage{graphicx}
\begin{document}

\title{Dendritic Actin Filament Nucleation Causes Traveling Waves and Patches}

\author{Anders~E~Carlsson}
\affiliation{Department of Physics, Washington University, One Brookings Drive, Campus Box
1105, St. Louis, MO~63130}

\date{\today}

\begin{abstract}
The polymerization of actin via branching at a cell membrane containing nucleation-promoting
factors (NPFs) is simulated using a stochastic-growth methodology.
The polymerized-actin distribution displays three types of behavior:
a) traveling waves, b) moving patches, and c) random fluctuations.
Increasing actin concentration causes a transition from patches to waves.
The waves and patches move by a treadmilling mechanism which does not require
myosin II. The effects of downregulation of key proteins 
on actin wave behavior are evaluated.
\end{abstract}

\maketitle

   The dynamic behavior  of the intracellular protein actin in cells is crucial in directing cell shape changes and migration. 
   Actin occurs in both monomeric  (``G-actin"), and polymerized (``F-actin") forms. F-actin consists of polar semiflexible
   filaments which grow preferentially at their ``barbed" ends.  The filaments assemble into
   supramolecular structures which provide forces to move cells or form dynamic protrusions along the cell periphery.
   A widely observed type of supramolecular structure is a branched dendritic network,
   where new filaments grow as branches on existing ones \cite{Mullins98,Pollard03}. 
   Several investigations have shown that F-actin in cells displays spontaneous dynamic behavior
   including traveling waves and patches \cite{Vicker02a,Vicker02b,Bretschneider04,Gerisch04}.
   The waves and patches in {\it Dictyostelium} typically move at about 0.2 $\mu m/s$.  A closely related 
   phenomenon is Hem-1 waves in neutrophils, where Hem-1 
   acts upstream of actin polymerization \citep{Weiner07}. Actin waves, when they impinge the membrane, cause 
   cellular protrusions to form \cite{Bretschneider09}. Therefore they may be involved in oscillatory and/or
   random motions of the cell membrane, which allow a cell to explore its environment. Thus understanding the origins of
   actin waves can help clarify the mechanisms by which cells determine key properties of their motion.

   The molecular mechanisms underlying actin patches and waves are not well known. However, a 
   partial molecular inventory of the waves has been obtained \cite{Bretschneider09}, including 
   Arp2/3 complex, which nucleates new filaments by branching, myosin-1B, which may link the
   actin network to the membrane, and coronin, which disassembles actin networks. 
   The presence of Arp2/3 complex suggests that patches could consist of dendritic actin networks;
   patches in yeast do in fact consist of such networks \citep{Young04}.

   Actin waves and patches might result \cite{LeGuyader97,Vicker02a,Doubrovinski08,Whitelam09}  from 
   a reaction-diffusion mechanism \citep{Mikhailov90}, based on spatially varying concentrations of an
   ``activator" and an ``inhibitor". 
    The activator grows at the front of the wave by positive feedback, and spreading (for example by diffusion)
    drives the wave forward.  The inhibitor then builds up, suppressing the activator at the rear of the wave.
   Actin waves and patches could contain positive feedback 
   either in F-actin (``actin-first"), in membrane-bound filament nucleation-promoting factors
   which act upstream of actin polymerization (``NPF-first"), or both. Ref. \citep{Whitelam09} treated a coarse-grained 
   model of F-actin dynamics with delayed membrane-induced inhibition
   and positive-feedback of F-actin on itself,
   which could implicitly contain contributions from NPFs.
   Assuming spontaneous polarization of actin filaments and
   diffusion-like spreading of F-actin led to the
   formation of patches and waves in appropriate parameter ranges.  Refs. \citep{Weiner07} and \citep{Doubrovinski08} 
   found spontaneous waves of F-actin and the NPF Hem-1 in models based on NPF cooperativity
   and inhibition of Hem-1 by F-actin;
   Ref. \citep{Doubrovinski08} also found actin patches under some conditions. 
   
   This Letter uses a detailed 3D dendritic network model to  treat 
   the spatial dynamics of F-actin, coupled to a NPF which shuttles
   between the cytoplasm and the membrane. This allows a treatment
   of actin waves and patches based on well-established molecular mechanisms,
   permits exploration of the possible dynamic phases of F-actin, and enables
   prediction of the dependence of F-actin dynamics on key biophysical rates and protein concentrations. 
   The main finding is that known molecular features of dendritic actin nucleation embody
   mechanisms of positive feedback, spreading, and delayed negative feedback which generate
   traveling waves and patches. This model differs from that of Ref. \cite{Whitelam09}, 
   in that polarization of filament orientations is not required. 
   Three broad types of F-actin dynamics are found: 1) traveling waves,
   2) moving patches, and 3) random fluctuations with occasional moving
   patches. 

The simulation method extends one used previously to treat
dendritic actin networks \citep{Carlsson01,Carlsson04a}. 
I treat network growth on a $3\mu m  \times 3\mu m $ piece of membrane. 
Explicit coordinates are stored for each filament. 
Polymerization, filament nucleation by branching, and capping, which
prevents barbed-end polymerization, are treated stochastically. 
New filaments are generated near the membrane as branches on the sides of
existing filaments, and by a slower non-branching nucleation process.
Branching is allowed only within 
10 nm  of the membrane, so that most branches form near
filament tips.
The method is extended by including the following effects  
(see Supplementary Material for more details):

{\it (i)} Filament severing, which is the primary actin disassembly mechanism in the
model. It is also an important actin filament nucleation mechanism \citep{Yamaguchi07}. 

{\it (ii)} Attractive filament-membrane forces, with binding energy $E_b$, 
required because the actin network is attached to the 
membrane \cite{Bretschneider09}.  

{\it (iii)} Translational and rotational motion of filaments, which are important for filament detachment
from the membrane and for ``docking" of filament clusters to the membrane. Because
this is very computationally demanding, 
I simplify the calculation by treating connected dendritic filament clusters 
as rigidly moving units.  
The translational cluster motion has a deterministic part from membrane forces,
and a random part from Brownian motion.

{\it (iv)} Membrane attachment, detachment, and diffusion of NPFs.
Detachment reduces or eliminates NPF activity, and 
F-actin enhances the detachment rate \cite{Weiner07}.
Taking the membrane as the x-y plane, the attached (active) NPF density $n_a (x,y)$ and spatially
averaged detached activator density ${\bar n}_d$ satisfy 
\begin{eqnarray}
{\partial n_a \over \partial t}&=&-k_{\rm det}F(x,y)n_a+k_{\rm att}{\bar n}_d+D_{\rm memb} \nabla^2 n_a  \nonumber \\
{d{\bar n}_d \over dt}&=&-\langle {\partial n_a \over \partial t}\rangle,
\end{eqnarray}
where $k_{\rm det}$, $k_{\rm att}$ are constants, $D_{\rm memb}$ is the attached NPF diffusion coefficient,
$\langle \rangle$ denotes 2D spatial averaging, and $F(x,y)$ is the 2D density of
F-actin (see Supplemental Material). Detached-NPF diffusion is assumed fast enough that $n_d(x,y)$ is constant. 
Because dimerization greatly accelerates the activity of attached NPFs \citep{Padrick08},  
I take the branching rate for a filament impinging the membrane at
$(x,y)$  to be proportional to $n_a(x,y)^2$.

The parameters (see Supplemental Material)  are taken from experiments where available and otherwise are chosen to obtain a 
realistic network structure. 
The on- and off-rates, the severing rate, the branch dissociation rate, and $D_{\rm memb}$
are obtained from experimental data. 
The value of [A] is varied within biologically reasonable limits, and the branching
rate and [CP] are chosen to give reasonable network structures;
$E_{\rm b}$, $k_{\rm att}$, $k_{\rm det}$, and the non-branching nucleation rate $k_{\rm nuc}$ are treated as independent variables.

\begin{figure}
   \begin{center}
    \includegraphics[width=12.5cm]{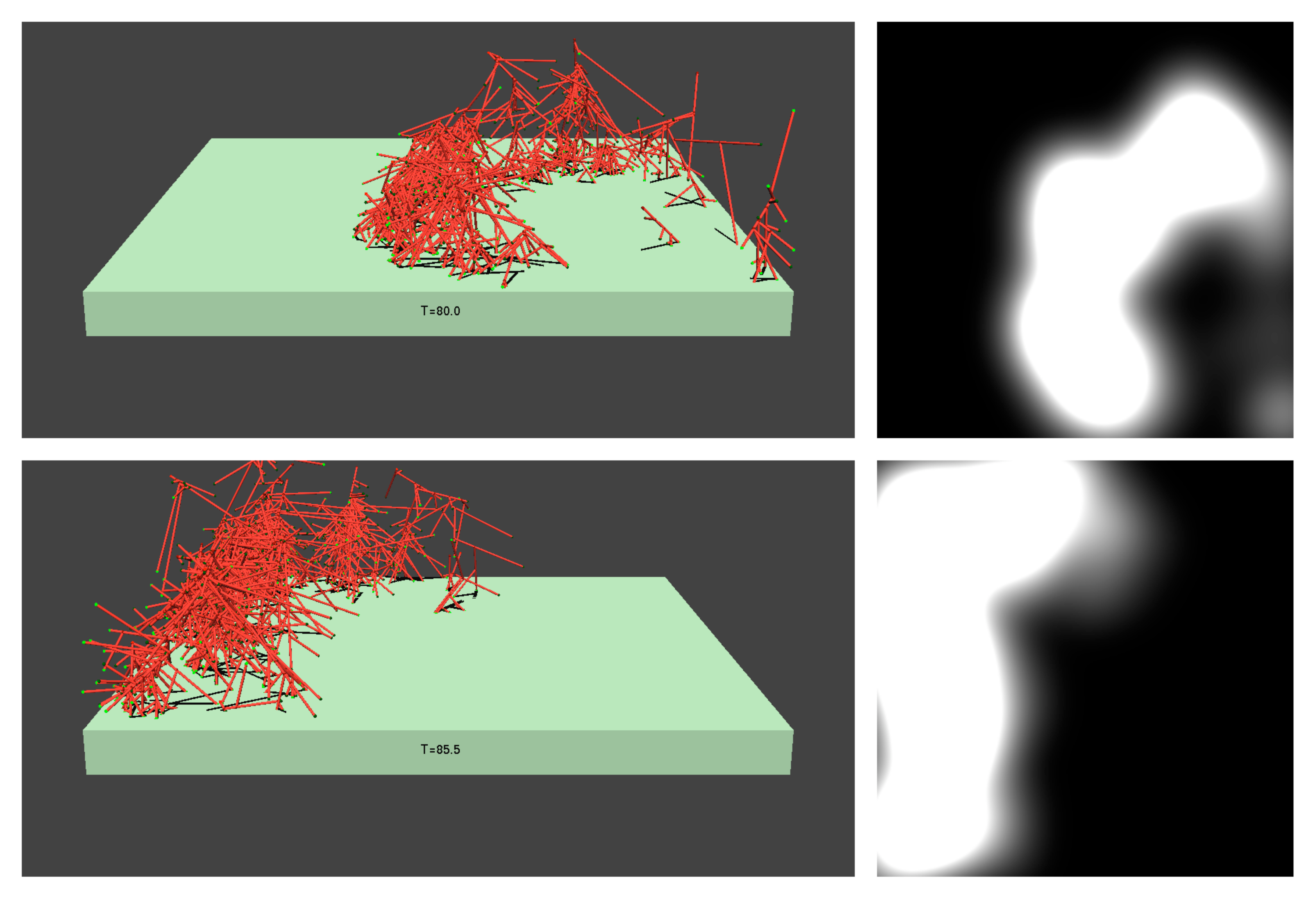}
\caption{3D network structure (left) and simulated fluorescence (right) for
actin wave (red) propagating across three-micron square area of membrane (green). 
Time difference between frames is 6 sec.
Parameters are: $k_{\rm att}=0.025s^{-1}$, $k_{\rm det}=0.009 \mu M^{-1}s^{-1}$, 
$E_{\rm b} =2.8k_{\rm B}T$, 
$k_{nuc} = 0.003 \mu M^{-1} s^{-1}$, 
$[A]=22.5 \mu M$, and $[CP]=0.25 \mu M$. See Supplemental Video 1 for complete time evolution.}
      \label{fig:waves}
   \end{center}
\end{figure}
The simulations begin with only G-actin present and are run out to 400s. Polymerization 
peaks and later reaches a steady-state dynamic regime, which is my focus. 
Fig. 1 shows snapshots of the 3D actin network and the simulated fluorescence intensity
(with a spread of 200 nm), under conditions favoring wave formation. 
I define a wave to be a moving feature having a large aspect ratio, while features with
smaller aspect ratios are patches; a quantitative description 
is given in the Supplementary Material.
A wave of polymerized actin moves across the membrane at a speed of about $0.2 \mu m/s$. The wave's shape 
changes as it moves, due to the randomness of polymerization and detachment. 
Although the F-actin features are mainly wavelike,
switches to a patchy state and back are sometimes seen in the simulations (see end of Video 1).

At lower [A], a patch initiates (left frame, Fig. 2)  as a localized focus of actin polymerization. 
It remains stationary for about 4s and then moves to the 
right at about $0.15 \mu m/s$, maintaining a roughly constant size.
The patch forms because G-actin becomes depleted 
before a complete wave can be formed. This result is consistent with previous studies of 
reaction-diffusion systems \citep{Krischer94}, which showed that global inhibition such as that mediated by
G-actin depletion causes transitions from waves to patches. 
The reduced [CP] pictures in Fig. 2 show a more random distribution of F-actin, 
with some patchy features which move and give the appearance of broken wave fronts.

\begin{figure}
   \begin{center}
    \includegraphics[width=8.5cm]{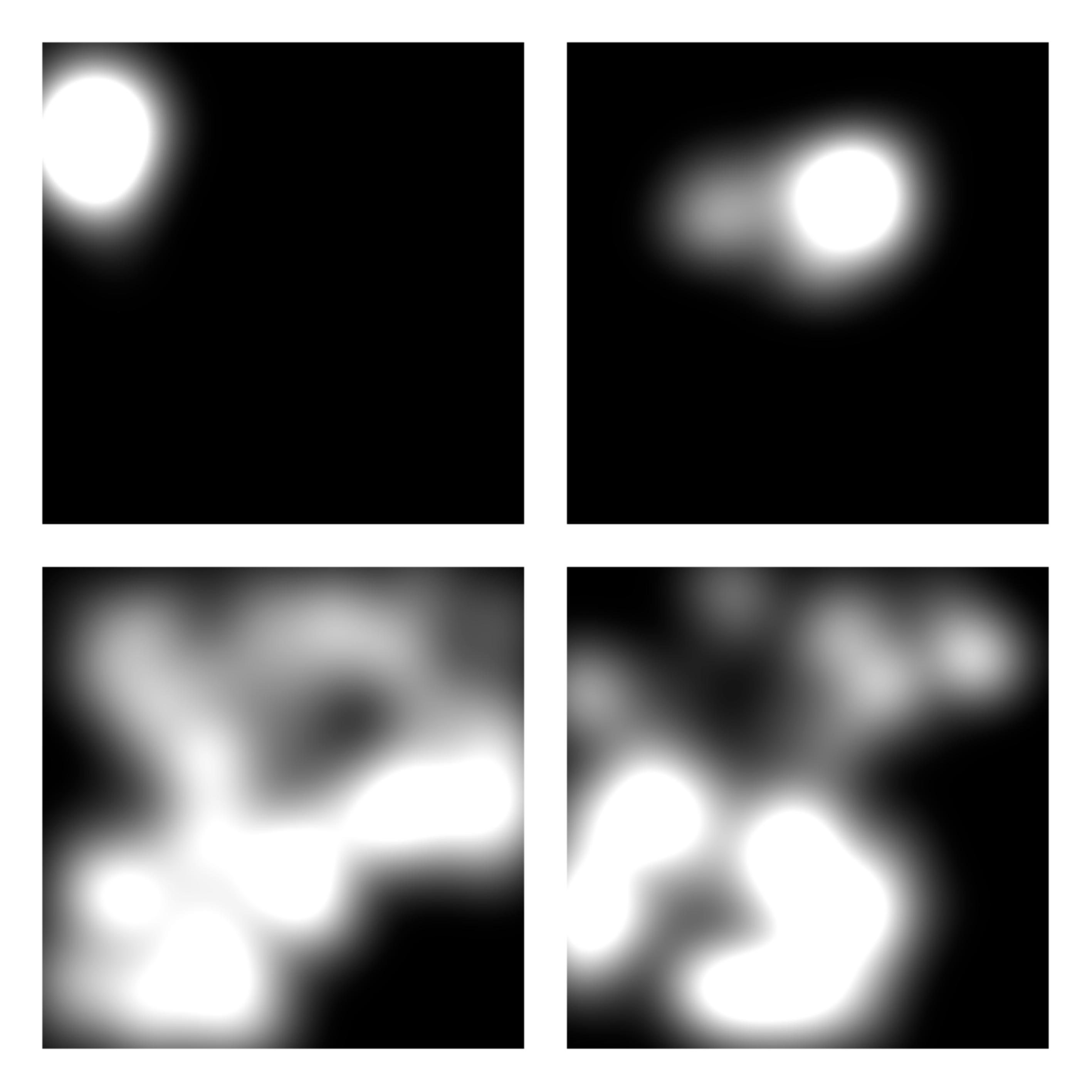}
\caption{Simulated fluorescence of patches obtained with parameters of Fig. 1 but 
with [A] reduced to $15 \mu M$ (top row) or [CP] decreased to $0.10\mu M$ (bottom row). 
Images are 12 sec (top row) and 18 sec (bottom row) apart.
See Supplemental Videos 2 and 3 for complete time evolution.}
      \label{fig:patches}
   \end{center}
\end{figure}

The motion of the waves and patches results from three factors operating jointly: 

{\bf Positive feedback of F-actin.} This is casued by
autocatalytic branching from existing filaments, which by itself
gives linear positive feedback \cite{Carlsson03} that is confirmed by {\it in vitro} polymerization dynamics\cite{Carlier00}. 
Branching is modulated
by filament detachment from the membrane.  If $E_{\rm b}$ is small, most single filaments leave the membrane before
they branch, so that formation of a dendritic cluster consisting of several filaments, 
by a statistical fluctuation, is required before continued growth occurs. The requirement for such a critical cluster
at small $E_{\rm b}$ values results in a sudden onset of polymerization, which favors wave/patch formation. 
This effect is similar to that of nonlinearity in the positive feedback.

{\bf Spreading of F-actin.} 
Fig. 3 shows a filament growing from a
patch by polymerization and nucleating several branches, thus providing new polymerization nucleui and spreading the patch
to the right.
This mechanism does not require the 
polarization of filament orientations suggested in Ref. \cite{Whitelam09},
since a broad range of filament orientations, including the vertical orientation, can generate new branches 
in the direction of motion (see Supplementary Material).
However, the patches and waves are polarized in that there are more free barbed ends at the leading edge.

{\bf Delayed negative feedback.} This effect is, as mentioned above, grounded in 
experimental observations. The delay has a large effect on the
F-actin dynamics when $1/k_{\rm att}$ exceeds the lifetime of polymerized actin, about
5 sec in the present simulations. 

These three factors can give rise to traveling waves and, when combined with 
global inhibition such as that arising from G-actin depletion, patches \cite{Krischer94}.
Small values of the non-branching nucleation rate $k_{\rm nuc}$ are also required in order to avoid ``parasitic"
nucleation interfering with wave or patch formation.

\begin{figure}
   \begin{center}
    \includegraphics[width=8.5cm]{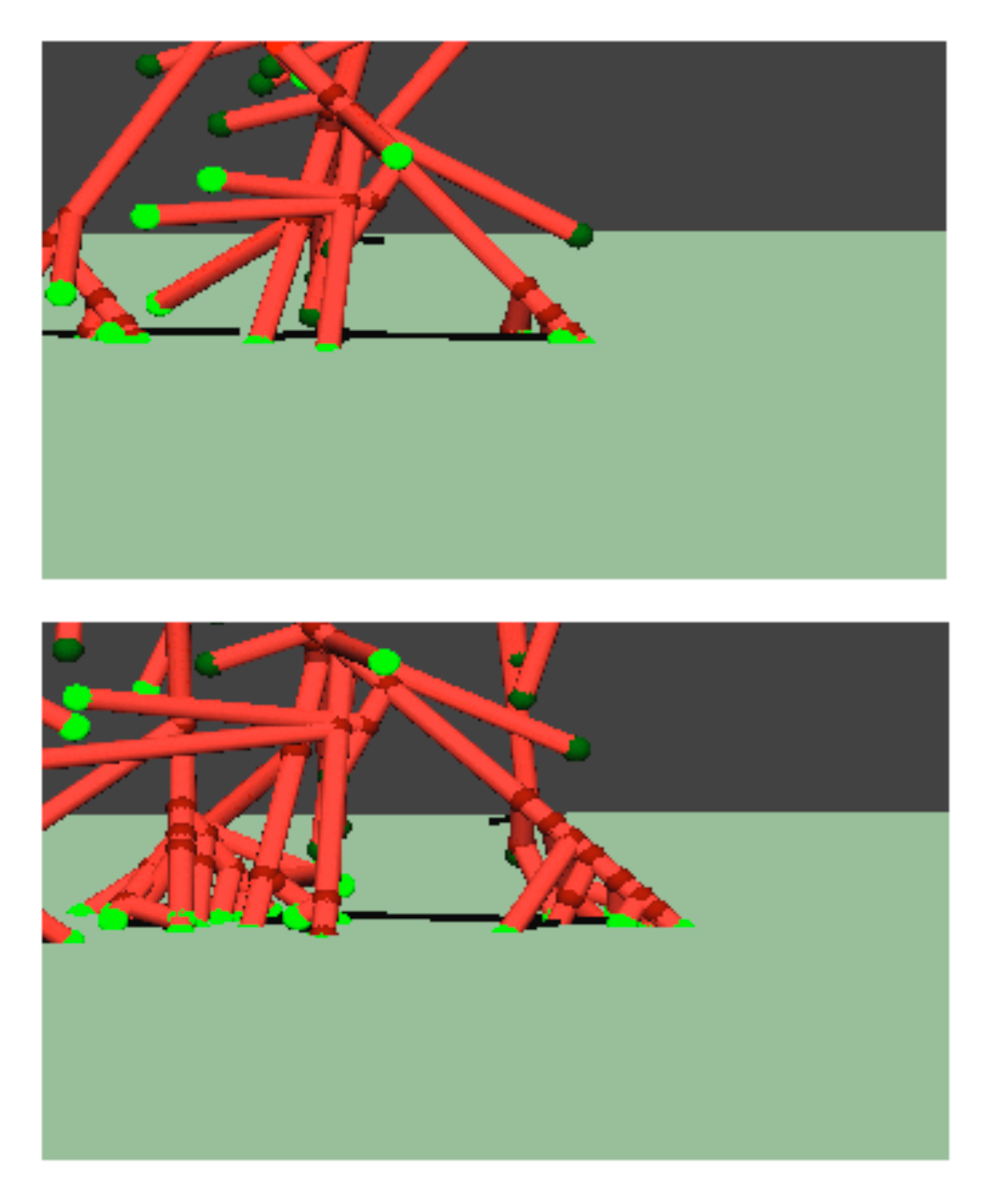}
\caption{Closeup of edge of actin patch at top of Fig. 2 moving to right. Frames are 0.2 sec apart.}
      \label{fig:cluster}
   \end{center}
\end{figure}

This model of actin waves and patches makes several experimentally testable predictions:

\noindent{1) Increasing G-actin concentration leads to a transition from stationary patches, to moving patches,
to waves (cf. Figs. 1 and 2). This prediction
is consistent with experiments  \cite{Gerisch04} that treated {\it Dictyostelium} cells with latrunculin,
which depolymerized actin by sequestering G-actin monomers. The latrunculin was then removed from the growth medium.
During the resulting recovery of polymerizable G-actin, the F-actin distributions
showed transitions from stationary patches, to moving patches, to waves. The patch-to-wave transition 
may well result from changes in [A], as in the present model. Because intracellular latrunculin is strongly bound
to G-actin, the latrunculin exit required for G-actin recovery could take minutes or longer.
The coexistence of waves and patches seen in Ref.  \cite{Gerisch04} is also consistent with 
the wave-patch switching seen in the simulations.}

\noindent{2) Colliding waves annihilate each other, since each depletes the NPF required for the other to persist. 
This is consistent with the experiments of Ref. \cite{Bretschneider09}}

\noindent{3)The waves and patches move by a treadmilling mechanism independent of myosin II (cf. Fig. 3).
This is consistent with photobleaching experiments \cite{Bretschneider09}, in which a bleached spot in the F-actin remained stationary
during wave motion and myosin II was not required for wave motion.}

\noindent{4) Arp2/3 complex is uniformly distributed throughout the wave (see Supplemental Material),
consistent with the fluorescence measurements of Ref. \cite{Bretschneider09}.}

\noindent{5) Uncapped barbed ends are polarized toward the front edge of patches and waves. This prediction
could be tested by standard techniques for measuring the spatial distribution of free barbed ends.}

\begin{figure}
   \begin{center}
    \includegraphics[width=11cm]{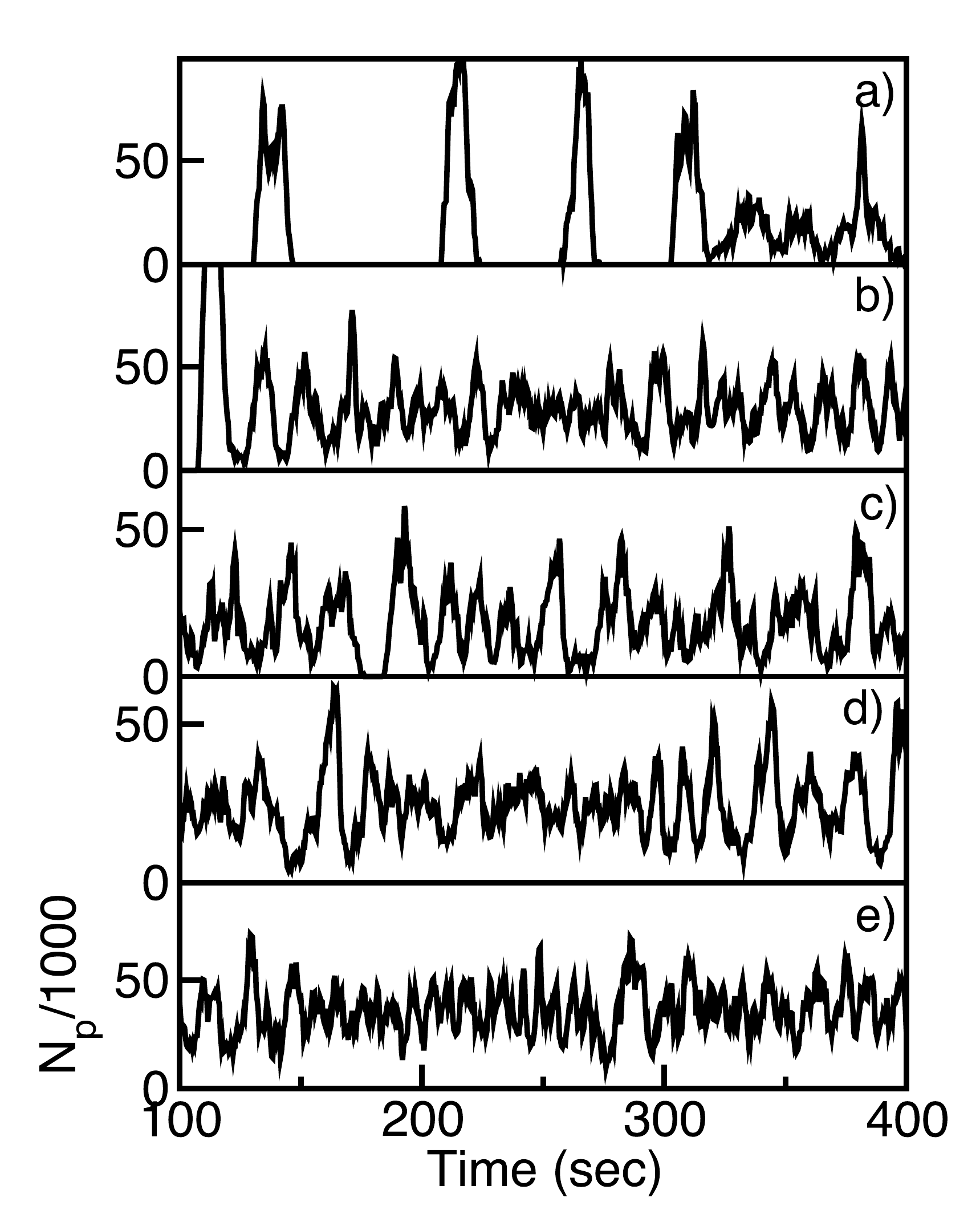}
\caption{Total number of polymerized subunits $N_{\rm p}$ vs time:
a) Baseline parameters as in Fig. 1;
b) [CP] decreased to $0.10\mu M$.;
c) $k_{nuc}$ increased to  $0.03 \mu M^{-1} s^{-1}$;
d) $E_{\rm b}$ increased to $4.5k_{\rm B}T$;
e) $k_{\rm att}$ increased to $0.05 s^{-1}$}
      \label{fig:npols}
    \end{center}
\end{figure}

\noindent{6) The actin filament orientations are isotropic rather than being biased along the direction of motion. This prediction could
be tested by cryo-electron microscopy measurements of wave or patch structure.}

\noindent{7) Downregulation of capping protein, increased nucleation activity in the membrane (as from formins), 
increased filament-membrane attachment strength (controlled by the protein VASP \cite{Samarin03}),
 or more rapid reattachment of NPFs to the membrane, will prevent wave or patch formation. Fig. 4 
shows the effects of such interventions on the dynamics of the number of polymerized subunits 
$N_{\rm p}$, used here as an approximate descriptor.
Large fluctuations in $N_{\rm p}$ (as in frame a) correspond to
traveling-wave or moving-patch states seen
in the simulation videos, while smaller fluctuations correspond to randomly fluctuating states.
All of the interventions reduce wave behavior. A recent study
\cite{Akin08} suggested that CP enhances branching {\it in vitro}, so that reduced CP levels might {\it reduce}
polymerization. However, in {\it Dictyostelium} reduced CP enhances polymerization \cite{Hug95}, consistent with
the assumptions made here. If CP enhances branching, the effects seen in Fig. 4b will be reduced. }

The main conclusion of this work is that essential mechanisms needed for formation of traveling waves and patches of F-actin follow from known properties of actin filament nucleation at membranes.  Thus the predictions are not limited to the particular mathematical model employed, but rather follow from the general mechanisms of F-actin positive feedback, F-actin spreading, and delayed negative feedback. The first two of these follow immediately from the nature of branching nucleation at membranes, while delayed negative feedback follows from plausible interactions mechanisms between F-actin and NPFs. In combination, these mechanisms lead to dynamic features closely paralleling those seen in experiments.

I appreciate informative discussions with Alexander Mikhailov, Martin Falcke, and Karsten Kruse. This work
was supported by the National Institutes of Health under Grant R01 GM086882.


\end{document}